\documentclass[pra,twocolumn,longbibliography]{revtex4}

\usepackage[utf8]{inputenc}
\usepackage{bm}
\usepackage{mathtools}
\usepackage{bbold}
\usepackage{amsmath}
\usepackage{amssymb}
\usepackage{hyperref}
\usepackage[normalem]{ulem}
\usepackage{color}
\usepackage{float}
\usepackage[dvipsnames]{xcolor}

\begin{document}

\title{Scalable spin squeezing in two-dimensional arrays of dipolar large-$S$ spins}
\author{Youssef Trifa and Tommaso Roscilde}
\affiliation{Univ Lyon, Ens de Lyon, CNRS, Laboratoire de Physique, F-69342 Lyon, France}


\begin{abstract}
Controlling the quantum many-body state of arrays of \emph{qudits}, possessing a large local Hilbert space, opens the path to a broad range of possibilities for many-particle entanglement, interesting both for fundamental quantum science, as well as for potential metrological applications. In this work we theoretically show that the spin-spin interactions realized in two-dimensional Mott insulators of large-spin magnetic atoms (such as Cr, Er or Dy) lead to scalable spin squeezing along the non-equilibrium unitary evolution initialized in a coherent spin state. An experimentally relevant perturbation to the collective squeezing dynamics is offered by a quadratic Zeeman shift, which leads instead to squeezing of individual spins. Making use of a truncated cumulant expansion for the quantum fluctuations of the spin array, we show that, for sufficiently small quadratic shifts, the spin squeezing dynamics is akin to that produced by the paradigmatic one-axis-twisting (OAT) model -- as expected from an effective separation between collective spin and spin-wave variables. Spin squeezing with OAT-like scaling is shown to be protected by the robustness of long-range ferromagnetic order to quadratic shifts in the equilibrium phase diagram of the system, that we reconstruct via quantum Monte Carlo and mean-field theory. 
\end{abstract}
\maketitle

\emph{Introduction.} The controlled production and certification of many-body entangled states \cite{Horodeckietal2009,Guehne_2009,TothA2014,Pezze2018RMP,Friisetal2019} is one of the most promising potentials of current quantum devices, based on \emph{e.g.} superconducting circuits \cite{Juanjobook} or on cold atomic ensembles \cite{Gross2017,BrowaeysL2020,Schaeferetal2020,Monroeetal2021,Kaufman2021NP,Chomazetal2022}. Most of the present platforms realize ensembles of interacting qubits, \emph{i.e.} $S=1/2$ spin systems, possessing the smallest possible local Hilbert space. Qubit systems can realize universal models of quantum computation \cite{IkeandMike}, and therefore already allow in principle for the realization of arbitrary quantum states for arbitrary degrees of freedom -- in which groups of qubits can be thought of realizing higher-dimensional objects. Yet working directly with \emph{qudits} \cite{Wangetal2020}, i.e. elementary degrees of freedom with a higher-dimensional Hilbert space, offers several advantages, both fundamental as well as practical. Systems of qudits are naturally realized in experiments using \emph{e.g.} photonic platforms \cite{Erhard2020}; molecular magnets \cite{Moreno-Pinedaetal2018}; and ensembles of large-$S$ magnetic atoms \cite{Chomazetal2022}. $N$ qudits can obviously encode an exponentially larger amount of quantum information than $N$ qubits;  entangled states of qudits can be more resilient to noise than entangled states of qubits \cite{Erhard2020}; and using qudits as quantum sensors \cite{Pezze2018RMP} instead of qubits can be very advantageous, in that single qudits already possess highly non-classical states with increased sensitivity to unitary transformations. The latter aspect also hints at a very intriguing competition that qudit systems (unlike qubit ones) can exhibit between single-qudit non-classical states and many-qudit non-classical (i.e. entangled) states. This competition will be a central aspect of the present work.

\begin{center}
\begin{figure*}[ht!]
\includegraphics[width=0.9\textwidth]{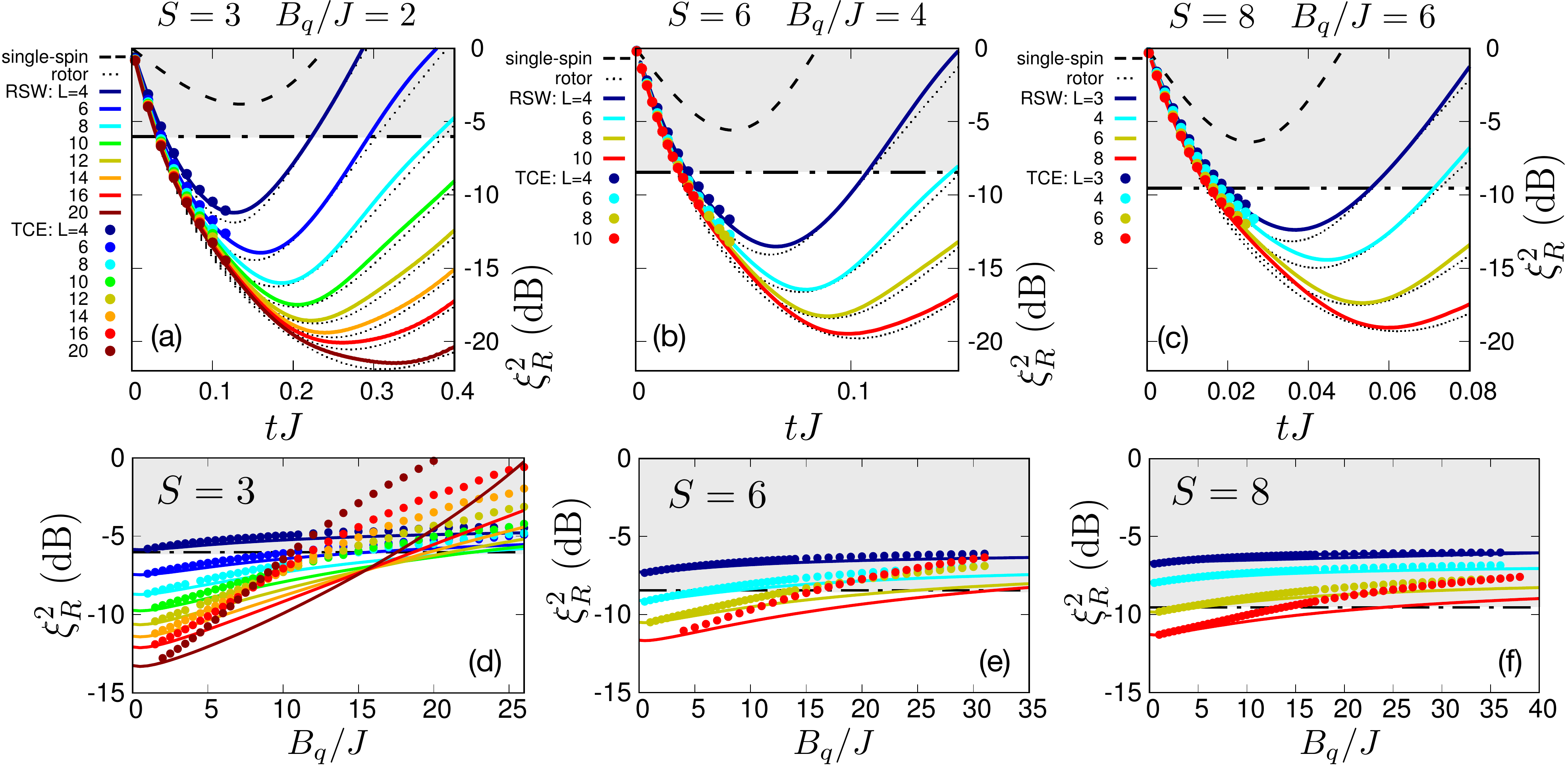}
\caption{\emph{Scalable spin squeezing in dipolar large-$S$ arrays.} (a-c) Time evolution of the squeezing parameter for various system sizes and $S=3$, $B_q/J=2$ (a), $S=6$, $B_q/J=4$ (b), and $S=8$, $B_q/J=6$ (a). Each panel shows the comparison between the single-spin limit; the rotor dynamics with effective moment of inertia $1/(2I)$ (see main text); the rotor/spin-wave theory (RSW) and the truncated cumulant expansion (TCE). The shaded region marks the regime $\xi_R^2 \geq (1+S)^{-1}$, in which squeezing does not witness entanglement. (d-f) Squeezing parameter at the time $0.3 t_{\rm min}$ (see main text) for various system sizes as a function of $B_q/J$ for $S=3$ (d), $S=6$ (e) and $S=8$ (f). Significance of symbols in panels (d), (e) and (f) is the same as in panels (a), (b) and (c), respectively.}
\label{f.OAT}
\end{figure*}
\end{center}

The focus of our work is the production of entangled many-qudit states in ensembles of large-$S$ magnetic atoms. Relevant examples -- that will be discussed later --  include $^{52}$Cr atoms ($S=3$), $^{168}$Er atoms ($S=6$) and  $^{162}$Dy atoms ($S=8$), whose spin degrees of freedom have been manipulated in a variety of recent experiments \cite{dePazetal2013,Lepoutreetal2019,Gabardosetal2020,Alaouietal2022,Patscheider2020PRR,Chalopinetal2018,Evrardetal2019,Satooretal2021}. 
When the above atoms are trapped in a deep optical lattice so as to form a Mott insulator with one atom per lattice site, their large spins interact at a distance via a dipolar interaction, whose Hamiltonian (in the rotating frame of an applied large Zeeman field) reads  \cite{Lepoutreetal2019,Patscheider2020PRR}
\begin{equation}
{\cal H} =   J \sum_{i<j} D_{ij} \left [ -\frac{1}{2}\left (S_i^x S_j^x +  S_i^y S_j^y \right ) + S_i^z S_j^z \right ] +  B_q \sum_i (S_i^z)^2 
\label{e.Ham}
\end{equation}
where $\sum_{i < j}$ runs over pairs of lattice sites; $S_i^\mu$ ($\mu = x,y,z$) are spin-$S$ operators; $J$ is the overall strength of the dipolar interaction; and $B_q$ is a quadratic Zeeman shift.  The factor $D_{ij} = (1-3\cos^2 \theta_{ij})/r_{ij}^3$ incorporates the distance and angle dependence of dipolar interactions, where  $\theta_{ij}$ is the angle formed between the applied Zeeman field and the vector ${\bm r}_{ij}$ connecting two sites $i$ and $j$. 
In the following we will imagine that the atoms are trapped in a square lattice, with a magnetic field perpendicular to the lattice plane, such that $\cos\theta_{ij} = 0$ for all pairs of spins. 
While recent experimental studies on dipolar atoms and molecules could detect the appearance of correlations induced by the dipolar interactions \cite{Alaouietal2022, Christakisetal2023}, the certification of entanglement in these systems is still lacking. 
 In this work we show that the dynamics of two-dimensional arrays of large-$S$ dipolar spins, initialized in a spin-coherent state in the $xy$ plane and governed by the Hamiltonian Eq.~\eqref{e.Ham}, can produce massive multipartite entanglement in the form of spin squeezing which is scalable, i.e. stronger the larger the number of atoms, following the paradigm of the one-axis-twisting (OAT) dynamics \cite{Kitagawa1993PRA}. 
 
 The collective OAT-like dynamics is robust to moderate values of the quadratic Zeeman shift $B_q$, since the Hamiltonian evolution develops long-range correlations in the $xy$ plane protecting the collective-spin length. At larger values of $B_q$ long-range order still persists in the thermalized state, but the total spin length is strongly reduced, so as to alter significantly the squeezing dynamics with respect to the OAT picture; or removing collective squeezing altogether in the dynamics of realistic system sizes.   
 The above results stem from a combination of different techniques, including dynamics based on a truncated-cumulant expansion of the quantum fluctuations in the evolved quantum state; a recently introduced rotor/spin-wave (RSW) separation approximation; and the reconstruction of the equilibrium phase diagram of the system. Our results pave the way for the use of large-spin dipolar arrays to produce metrologically useful, multipartite entangled states.

\begin{center}
\begin{figure*}[ht!]
\includegraphics[width=\textwidth]{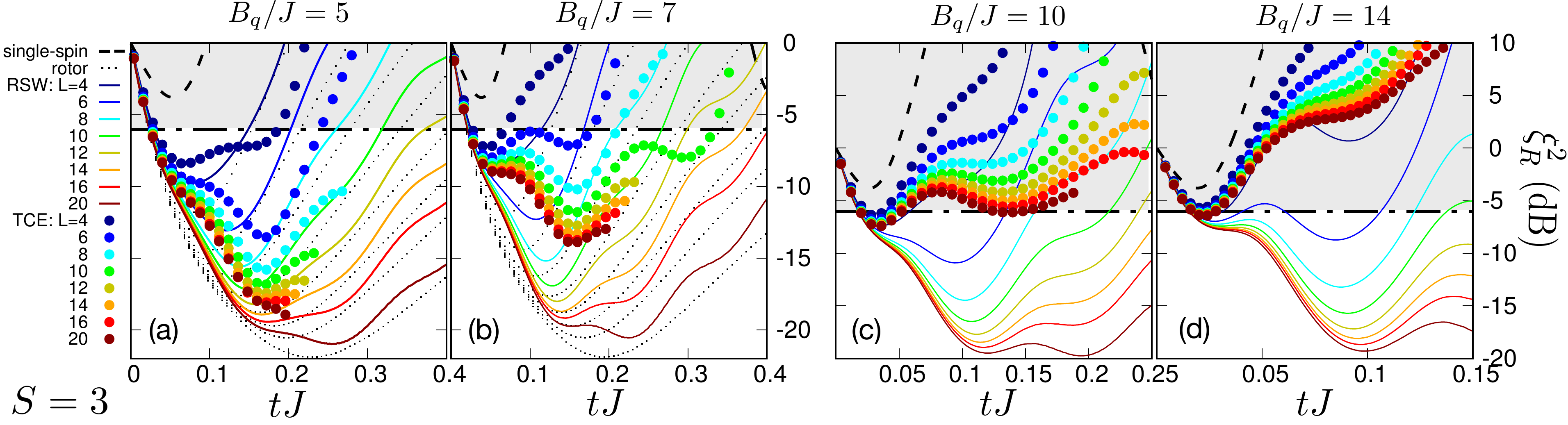}
\caption{\emph{Squeezing dynamics at large $B_q$.} Time evolution of the spin squeezing parameter for $S=3$ and three values of $B_q/J=$ (a) 5, (b) 7, (c) 10 and (d) 14. All symbols are as in Fig.~\ref{f.OAT}(a).}
\label{f.variousBq}
\end{figure*}
\end{center}

\emph{Spin squeezing and entanglement; squeezing dynamics from RSW separation.} 
The paradigmatic example of many-body dynamics giving rise to scalable squeezing is offered by the one-axis-twisting (OAT) model, whose Hamiltonian has the form of a planar rotor whose angular momentum along the $z$ axis is given by the collective spin, namely ${\cal H}_{\rm OAT} = (J^z)^2/(2I)$. Here we have introduced the collective spin operator ${\bm J} = \sum_{i=1}^N {\bm S}_i$ for the ensemble of $N$ quantum spins; and $I \sim {\cal O}(N)$ is the extensive moment of inertia of the rotor. When the system of $N$ spins of length $S$ is initialized in a coherent spin state (CSS) along $x$,  $|{\rm CSS}_x\rangle = |S;x\rangle^{\otimes^N}$ (where $S^x |S;x\rangle = S |S;x\rangle$), the collective spin is of maximal length ${\bm J}^2 = J_{\rm max}(J_{\rm max}+1)$ with $J_{\rm max} = NS$ .  The Hamiltonian evolution governed by ${\cal H}_{\rm OAT}$ remains in the $J_{\rm max}$ sector, and it produces squeezing of the collective spin, captured by the squeezing parameter 
\begin{equation}
\xi_R^2 = \frac{2NS~{\rm Var}(J^{\rm min})}{\langle J^x \rangle^2}
\label{e.xiR2}
\end{equation}  
where ${\rm Var}(J^{\rm min}) = \min_{\perp} {\rm Var}(J_{\perp})$ is the minimal variance of the collective spin components in the $yz$ plane, transverse to the average spin orientation (along $x$). 
The initial CSS has $\xi_R^2=1$, while $\xi_R^2<1$ indicates squeezing of the uncertainty along one direction in the $yz$ plane with respect to this reference state. The OAT dynamics is known to produce an optimal (scalable) squeezing $(\xi_R^2)_{\rm min} \sim (2NS)^{-2/3}$ at a time $t_{\rm min} \sim S^{-2/3} N^{1/3}$.
Yet, for $S>1/2$ the condition  $\xi_R^2<1$  can also be satisfied at the level of individual spins -- as \emph{e.g.} generated by a collection of single-spin OAT models, ${\cal H}_{\rm 1s-OAT} = B_q \sum_i (S_i^z)^2$.  Therefore spin squeezing per se is not necessarily a witness of entanglement. Nonetheless, uncorrelated $S$-spins can achieve a minimum squeezing parameter given by  $(\xi_R^2)_{\rm SQL} = (1+S)^{-1}$ (for $S>1/2$) \cite{Pezze2018RMP}, corresponding to the standard quantum limit (SQL) of squeezing for large-$S$ spins. Hence the condition $\xi_R^2  < (1+S)^{-1}$  certifies the presence of entanglement; and the stronger condition  $\xi_R^2  < [(1+kS)]^{-1}$ signals $(k+1)$-partite entanglement.  

The OAT model offers an important paradigm for the squeezing dynamics of dipolar Hamiltonians in 2$d$ such as Eq.~\eqref{e.Ham} -- as shown in recent numerical as well as experimental works, \cite{Perlin2020PRL,Comparinetal2022, Comparinetal2022b,Blocketal2023, Bornetetal2023}. Indeed, while the dipolar dynamics does not conserve the collective spin length ${\bm J}^2$, it may lead to a moderate decrease with respect to its maximum value, which justifies a scenario of rotor/spin-wave (RSW) separation, as proposed in Refs.~\cite{Roscildeetal2023, Roscildeetal2023b}. In a nutshell, the dipolar Hamiltonian, when projected onto the $J_{\rm max}$ sector of symmetric states, takes the form of a  OAT Hamiltonian with effective moment of inertia $1/(2I) = \frac{J}{2N^2} \sum_{i\neq j} D_{ij}  + \frac{B_q}{N}$, which would govern the dynamics if restricted to the $J_{\rm max}$ sector. Leakage out of this sector can be accounted for by the production of spin-wave excitations, which can be described as finite-momentum Holstein-Primakoff (HP) bosons. If such bosons form a dilute gas, they can be described as a system of free quasiparticles; and they can be considered as effectively decoupled from the collective spin projected on the $J_{\rm max}$ sector (hereafter called rotor). The dynamics of the system can therefore be cast as the independent dynamics of a rotor variable $\bm K$ of length $K= J_{\rm max}$, governed by the OAT Hamiltonian ${\cal H}_R = (K^z)^2/(2I)$; and of linearized spin waves at finite momentum ${\cal H}_{\rm sw} = \sum_{\bm k \neq 0} (b^\dagger_{\bm k}, b_{-\bm k}) h_{\bm k} (b_{\bm k}, b^\dagger_{-\bm k})^T$, where $h_{\bm k}$ is a 2x2 matrix and $b_{\bm k}$ are HP boson operators onto which the spins have been mapped (see Supplemental Material -- SM \cite{SM} -- for further details). 
Within RSW theory \cite{Roscildeetal2023, Roscildeetal2023b}, the squeezing parameter is expressed as $\xi_R^2 \approx (2NS)~ {\rm Var}(K^{\rm min})/(\langle K^x \rangle - N_{\rm bos})^2$, where  $N_{\rm bos} = \sum_{\bm k \neq 0} \langle b_{\bm k}^\dagger b_{\bm k}\rangle$ is the population of HP bosons. 
Hence the squeezing dynamics of the system is akin to that of the OAT model (and showing the same scaling behavior) so long as $N_{\rm bos}$ remains a small correction to the rotor magnetization up to the time $t_{\rm min}$. {As we shall detail in the SM \cite{SM}, the spin-wave dynamics develops instabilities (due to imaginary frequencies) for negative values of $B_q$, due to a ground-state phase transition occurring for $B_q \lesssim -J$ (for all the values of $S$ we explored) -- see further details below; therefore we shall focus on positive $B_q$ values in the following.}

\emph{OAT-like regime and its breakdown.} Figs.~\ref{f.OAT}(a-c) shows indeed that, for $B_q>0$, the picture offered by RSW theory applies to the dynamics generated by the Hamiltonian Eq.~\eqref{e.Ham} as long as $B_q$ is sufficiently small. The predictions of RSW theory are indeed found to only moderately deviate from those of a pure-rotor dynamics for variable $B_q$ ranges, which grow with the growth of the spin length $S$ -- this means that spin waves are dilute, justifying the separation of variables. Most importantly, at short times the RSW predictions are corroborated by a completely alternative approach based on a truncated cumulant expansion (TCE) -- up to 2nd order cumulants -- for the multivariate quantum fluctuations of the spin ensemble. The TCE approach (described in details in the SM \cite{SM}) is a variant of similar approaches developed for quantum many-body systems and in quantum optics \cite{Rudiger1990,Leymann2014,Plankensteiner21,Colussi2020,Verstraelenetal2023}; the specificity of our formulation for large-$S$ systems is that it can describe exactly the physics at the single-spin level, in spite of the large local Hilbert space -- a trait which is essential when single-spin physics competes with many-body physics, as we shall see later. The assumption of truncation of the cumulant hierarchy leads to unphysical results after a given time, restricting this approach to short times when strong squeezing driven by the many-body physics appears.  

An extensive comparison of the predictions of the RSW and TCE approaches for the scaling of squeezing is offered in Fig.~\ref{f.OAT}(d-f). There we plot the value of the squeezing parameter at a time which does not correspond to its absolute minimum, but rather to an earlier time, due to the breakdown of the TCE approach mentioned above. We choose the time of observation as  $0.3 t_{\rm min}$, where $t_{\rm min}$ is the optimal time at which the OAT model with coupling constant $1/(2I)$ reaches its minimum squeezing. As we show in the SM \cite{SM}, scalable squeezing is expected in the OAT model at the optimal time as well as at earlier times, albeit with a slower scaling than at the optimal time. As one can see, RSW and TCE results agree well for all system sizes over the ranges $B_q \lesssim  2J$ for $S=3$; $B_q \lesssim  5J$ for $S=6$; and $B_q \lesssim  7J$ for $S=8$. These ranges correspond therefore to OAT-like scalable squeezing. For larger $B_q$ values, the two theories (TCE and RSW) start to deviate, signaling that the picture of separation of variables underlying RSW theory breaks down. Nonetheless, as we shall also discuss in the next section, squeezing appears to remain scalable for a larger range of $B_q$ values, albeit following a behavior which can no longer be understood starting from the OAT paradigm. 

\begin{center}
\begin{figure}[ht!]
\includegraphics[width=\columnwidth]{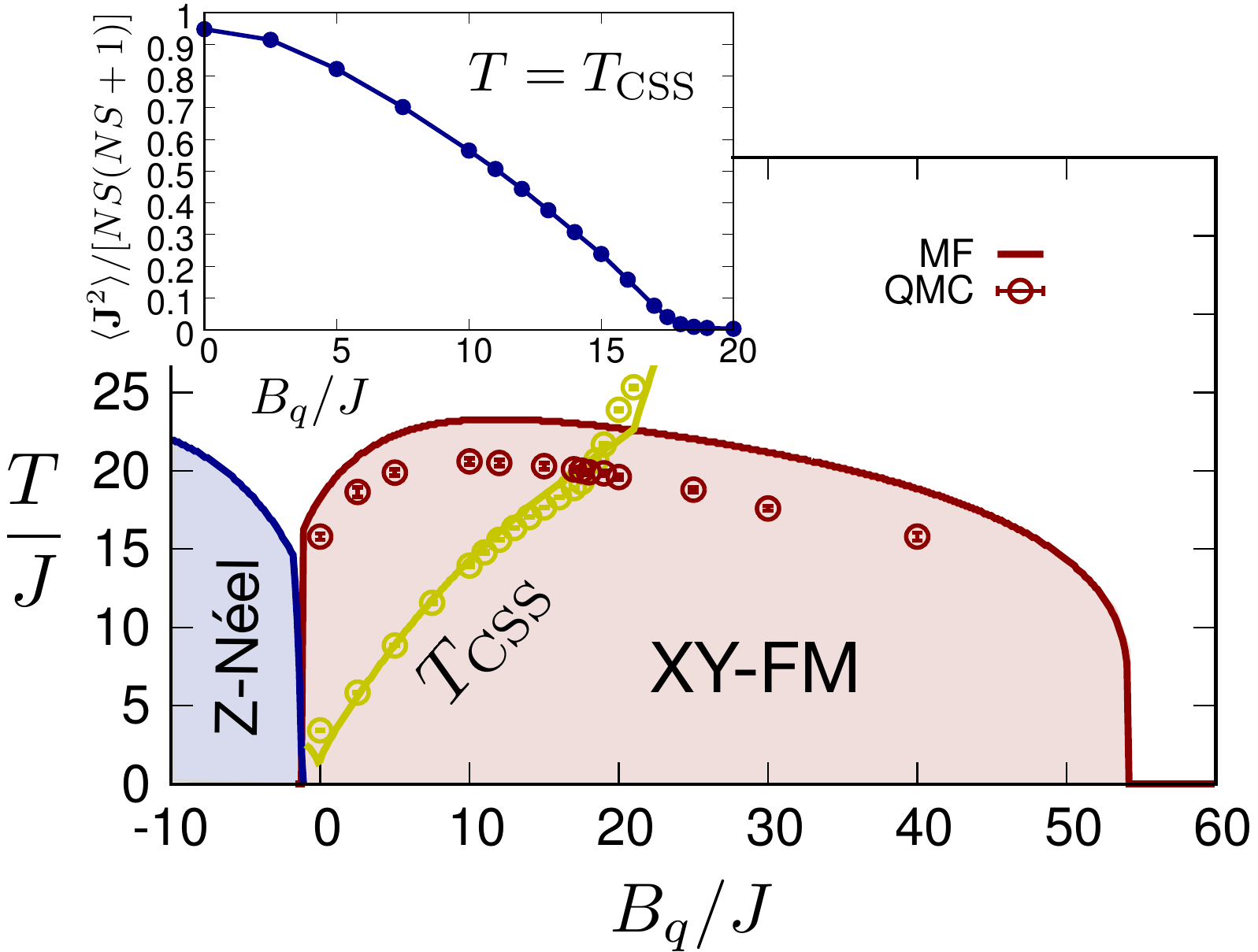}
\caption{\emph{Equilibrium phase diagram of the $S=3$ dipolar XXZ model on the square lattice.} The diagram shows the mean-field (MF) and quantum Monte Carlo (QMC) estimates of the transition temperature to $xy$-ferromagnetism (XY-FM), to N\'eel antiferromagnetism (Z-N\'eel), and of the coherent-state temperature $T_{\rm CSS}$. Inset: collective-spin square modulus $\langle {\bm J}^2 \rangle$ along the $(B_q, T_{\rm CSS}(B_q))$ line, evaluated for a system of $N = 48\times 48$ spins.}
\label{f.thermodynamics}
\end{figure}
\end{center}

\emph{Scalable vs. non-scalable spin squeezing, and relationship to thermodynamics.} To understand the evolution in the scaling of the squeezing parameter upon increasing $B_q$, it is useful to consider the limit of $B_q \gg J$, in which the dynamics is dominated by the quadratic Zeeman shift. In this limit (denoted by the dashed curves in Figs.~\ref{f.OAT} and \ref{f.variousBq}) each spin behaves as an OAT model, exhibiting a fast depolarization over a time $\sim B_q \sqrt{S}$ which is system-size independent -- this is the dynamics realized in recent experiments on single Dy atoms in Ref.~\cite{Chalopinetal2018}. How is this single-spin dynamics connected with the collective-squeezing dynamics at small $B_q$? 
Fig.~\ref{f.variousBq} shows that when leaving the OAT-like regime, scalable squeezing persists, albeit with very large fluctuations in time. For sufficiently small $B_q$ these fluctuations are partly reproduced by RSW theory (Fig.~\ref{f.variousBq}(a)), which attributes them to spin waves -- although the proliferation of finite-momentum spin waves is clearly at the core of the breakdown of the separation-of-variable assumption underlying RSW theory \cite{SM}. 
As $B_q$ increases, the competition between single-spin and many-body physics becomes manifest (see e.g. Figs.~\ref{f.variousBq}(c-d)), with a short-time behavior, dominated by the $B_q$ term, in which $\xi_R^2$ reaches a nearly size-independent first minimum, followed by anti-squeezing dynamics ($\xi_R^2$ increases). This first phase is followed by a second phase in which, at least for the largest system sizes, anti-squeezing stops, and $\xi_R^2$ begins to decrease again developing a second minimum, the deeper the larger the size. This later dynamics is clearly the result of many-body physics -- as revealed by its scaling nature -- and it can be understood in relationship to a fundamental trait of dipolar spins in 2d, namely their ability to develop long-range order at low energy \cite{Peteretal2012}. 

At this point it is instructive to inspect the equilibrium phase diagram of the dipolar large-$S$ Hamiltonian Eq.~\eqref{e.Ham} in 2d, which we have reconstructed in the temperature-vs-$B_q$ plane using numerically exact quantum Monte Carlo, as well as mean-field theory. The results are shown in Fig.~\ref{f.thermodynamics} for the case $S=3$ (see SM \cite{SM} for analogous phase diagrams for $S=6$ and 8, and details about the calculations): for a large $B_q$ region, $0 \lesssim B_q \lesssim 54 J$, the thermodynamics exhibits long-range ferromagnetism in the $xy$ plane up to a critical temperature $T_c$ which reaches a peak value of $\approx 20 J$; while for $B_q \lesssim -1.1 J$ the system exhibits long-range N\'eel antiferromagnetism along the $z$ axis. 
When initialized in the CSS, the unitary evolution driven by the Hamiltonian Eq.~\eqref{e.Ham} at long times is expected to thermalize the state of the system \footnote{Albeit the thermalization process can be highly non-standard, with the appearance of highly non-thermal states such as Schr\"odinger's cat states, as highlighted in Ref.~\cite{Comparinetal2022b}.}, so that the time average of local observables reproduce their equilibrium expectation values at a temperature $T_{\rm CSS}$ such that $\langle {\rm CSS}_x | {\cal H} | {\rm CSS}_x \rangle = \langle {\cal H} \rangle_{T_{\rm CSS}}$, where $\langle ... \rangle_T$ denotes the thermodynamic average at temperature $T$. The unitary evolution at long times is therefore sensitive to the equilibrium behavior along the $T_{\rm CSS}(B_q)$ line in the phase diagram (see Fig.~\ref{f.thermodynamics}) along which the state transitions from $xy$-ferromagnetic (XY-FM) to paramagnetic for $B_{q,c} \approx 18 J$. In particular this transition is marked by a drop of the collective spin length $\langle \bm J^2 \rangle$ from macroscopic values ($\sim O(N^2)$) to microscopic values ($\sim O(N)$) -- see inset of Fig.~\ref{f.thermodynamics}. 

Spontaneous breaking of the $U(1)$ symmetry of Eq.~\eqref{e.Ham}  in the thermodynamic limit can have important consequences on the squeezing dynamics \cite{Blocketal2023}, as it implies that, on finite system sizes,  the initial polarization $\langle J^x \rangle$ associated with the CSS persists for increasingly long times the larger the system size (and never dies out in the thermodynamic limit). This is shown by our data (see SM \cite{SM}), and as also observed experimentally in systems of dipolar qubits \cite{Bornetetal2023}. As a consequence, in spite of the depolarizing effect coming from the $B_q$ term, ferromagnetism can protect the spin-squeezing parameter from blowing up, because it prevents a fast vanishing of the denominator in Eq.~\eqref{e.xiR2}; and it can delay or even reverse the anti-squeezing dynamics. Moreover the collective spin remains of macroscopic length, as guaranteed by the fact that $\langle \bm J^2\rangle \sim O(N^2)$ throughout the evolution -- although it may depart significantly from its (initial) maximum value $NS(NS+1)$, as shown in Fig.~\ref{f.thermodynamics}. 

Nonetheless, as we clearly observe in Fig.~\ref{f.variousBq}(c,d) (see also the SM \cite{SM} for further data), for the system sizes we explored (up to $N=400$ for $S=3$) the fast depolarization imposed by the $B_q$ term can push the squeezing parameter to values $\xi_R^2$ which are systematically higher than the entanglement threshold $\xi_R^2 = (1+S)^{-1}$ (for $B_q \gtrsim 10 J$), or even higher than the proper squeezing threshold $\xi_R^2 = 1$ (for $B_q/J \gtrsim 13 J$ \cite{SM}), so that collective spin squeezing (namely squeezing exceeding what can be achieved at the level of single spins) is lost for $B_q$ values well below $B_{q,c}$.  
This situation may however coexist with a persistent scaling of the squeezing parameter to lower values -- namely with persistent scalable squeezing; nonetheless extremely large system sizes, beyond the reach of current experimental setups, may be required to bring $\xi_R^2$ to values which are compatible with entanglement, and which therefore offer many-body metrological advantage compared to single spins (see Ref.~\cite{Chalopinetal2018}). Long-range ferromagnetism in the thermalized state is only a necessary condition for scalable squeezing \footnote{Indeed scalable squeezing is not present in the thermalized state at any finite temperature \cite{Comparinetal2022}.}; for scalable squeezing to appear at intermediate times not only should $\langle J^x \rangle/N$ not scale to zero, but also ${\rm Var}(J^{\rm min})/N$ must scale to ever lower values with increasing size. Such a behavior is apparent in our data \cite{SM}, and even persisting for $B_q>B_{q,c}$, although a power-law decay of  ${\rm Var}(J^{\rm min})/N$ with $N$ is not revealed by our data. {Therefore our results are not inconsistent the conjecture of Ref.~\cite{Blocketal2023} that scalable squeezing persists up to the transition in the thermalized state; yet, for the system sizes we explored, the scaling behavior close to the transition appears to be very different from (and much slower than) that of the OAT model.}

\emph{Conclusions.}  In this work we have shown that the non-equilibrium dynamics of 2d arrays of dipolar large-$S$ spins, initialized in a coherent spin state, features multipartite entanglement in the form of scalable spin squeezing, obeying the scaling of the one-axis-twisting model for a sufficiently small quadratic Zeeman shift, in agreement with a scenario of 
separation of variables between collective-spin and spin-wave degrees of freedom.
Our results point at the crucial role played by the quadratic Zeeman shift on the squeezing dynamics of large-$S$ spins -- and, more generally, at the competition between single-qudit vs. many-qudit Hamiltonian in the entangling dynamics of qudit ensembles. In order to achieve collective spin squeezing in 2d dipolar arrays, the quadratic Zeeman shift should be controlled to the level of $\sim 10 J$ (with $J$ the spin-spin interaction energy), which can be achieved via magnetic fields and tensor light shifts \cite{Patscheider2020PRR, Chalopinetal2018}. The two-dimensional geometries we explored in this work are essential for spin-squeezing dynamics to occur: indeed, due to its angular dependence the dipolar interaction averages to zero in three dimensions, so that the collective-spin dynamics is suppressed in 3d. Nonetheless purely 2d arrays of atoms can be realized either by loading a single layer in a three-dimensional optical lattice, or by trapping in quantum-gas-microscope setups, as recently demonstrated experimentally for dipolar Er atoms \cite{Suetal2023}. Hence our work paves the way for the  realization of scalable multipartite entanglement in arrays of magnetic atoms (Cr, Er or Dy), representing a most promising platform to realize quantum simulation and quantum information processing with ensembles of qudits.

\begin{acknowledgements}
 \emph{Acknowledgements.} This work is supported by ANR (EELS project) and QuantERA (MAQS project). Fruitful discussions with M. Block, N. Defenu, B. Laburthe, L. Vernac, N. Yao, and B. Ye are gratefully acknowledged. All numerical simulations have been performed on the PSMN cluster at the ENS of Lyon.   
\end{acknowledgements}

\newpage

\noindent
{\bf Supplemental Material} \\
\noindent
{\bf \emph{
Scalable spin squeezing in two-dimensional arrays of dipolar large-$S$ spins}}

\section{Rotor/spin-wave theory for large-$S$ spins} 

In this section we review the basic elements of rotor/spin-wave (RSW) theory  \cite{Roscildeetal2023, Roscildeetal2023b} as applied to the XXZ Hamiltonian (Eq.~(1) of the main text) for large-$S$ spins. 

RSW theory for a U(1) symmetric system assumes that the ground state develops long-range order in the $xy$ plane. After mapping the spin operators onto bosonic ones $b_i$, $b_i^\dagger$ (and $n_i = b_i^\dagger b_i$) via the Holstein-Primakoff (HP) transformation
\begin{eqnarray}
 S_i^x &=&  S - n_i \nonumber \\
 S^y_i &=& \frac{1}{2} (\sqrt{2S-n_i} ~b_i + {\rm h.c.}) \nonumber \\
 S^z_i &=& \frac{1}{2i} (\sqrt{2S-n_i} ~b_i - {\rm h.c.}) ~
 \label{e.HP}
 \end{eqnarray}
the Hamiltonian of the XXZ model can be recast as a bosonic Hamiltonian containing nonlinear terms up to infinite order. Moving to bosonic operators in momentum space
\begin{equation}
b_{\bm k} = \frac{1}{\sqrt{N}} \sum_i e^{-i {\bm k} \cdot \bm r_i} b_i
\end{equation}
and reorganizing the bosonic Hamiltonian so as to regroup all the terms containing \emph{exclusively} the zero-momentum operators $b_0$, $b_0^\dagger$, leads to the (approximate) RSW decomposition  
\begin{equation}
{\cal H}_{\rm XXZ} = {\cal H}_{\rm R} + {\cal H}_{\rm SW} + \text{(coupling terms)}
\label{e.RSW}
\end{equation} 
containing the following terms:
\begin{itemize}
\item the rotor Hamiltonian
\begin{equation}
{\cal H}_{\rm R} = E_{0,\rm R} + \frac{(K^z)^2}{2I} 
\end{equation}
describing a macroscopic spin variable $\bm K$ of length $\bm K^2 = K(K+1)$ with $K = NS$ built out of the $b_0, b_0^\dagger$ operators,
$K^x = NS - n_0$,  $K^y = \frac{1}{2} (\sqrt{2NS-n_0} ~b_0 + {\rm h.c.})$  and $K^z = \frac{1}{2i} (\sqrt{2NS-n_0} ~b_0 - {\rm h.c.})$ with $n_0 = b_0^\dagger b_0$, whose inverse moment of inertia reads
\begin{equation}
\frac{1}{2I} = \frac{J}{2N^2} \sum_{i\neq j} D_{ij}  + \frac{B_q}{N}~.
\label{e.inertia}
\end{equation}
$E_{0,\rm R} =  -\frac{S(NS+1)}{2N} \sum_{i\neq j} D_{ij}$~is the rotor zero-point energy;

\item the spin-wave Hamiltonian, collecting all the quadratic terms involving finite-momentum bosonic operators $b_{\bm k \neq 0},  b^\dagger_{\bm k \neq 0}$
\begin{equation}
{\cal H}_{\rm SW} = \sum_{\bm k \neq 0} \left [{\cal  A}_{\bm k} b_{\bm k}^\dagger b_{\bm k} + \frac{{\cal B}_{\bm k}}{2} \left ( b_{\bm k} b_{-\bm k} + {\rm h.c.} \right ) \right ] 
\label{e.SMHam}
\end{equation}
where 
\begin{eqnarray}
{\cal A}_{\bm k} & = &  S \left ( J_0 + J_{\bm k}/2 + B_q \right)   \nonumber \\
{\cal B}_{\bm k} & =  & - S \left ( 3 J_{\bm k}/2~ +B_q \right ).
\label{e.AkBk}
\end{eqnarray}
Here we have introduced the Fourier transform of the dipolar coupling
\begin{equation}
J_{\bm k} = \frac{J}{N} \sum_{i\neq j} e^{i{\bm k} \cdot (\bm r_i - \bm r_j)} ~D_{ij} ~.
\end{equation}

\item the residual coupling terms, which couple the zero-momentum bosons with the finite-momentum ones. If the Hamiltonian is defined on a translationally invariant lattice, as it is the case throughout this work, the lowest-order coupling terms are quartic in the bosonic operators, namely they are parametrically smaller than the rotor and spin-wave Hamiltonians as long as the number of HP bosons remains small, and particularly so the number of finite-momentum bosons, $N_{\rm bos} = \sum_{\bm k \neq 0} \langle b_{\bm k}^\dagger b_{\bm k} \rangle$.    

\end{itemize}

RSW theory amounts to neglecting the coupling terms in Eq.~\eqref{e.RSW}, resulting in an effective separation of variables between the (zero-momentum) rotor and the (finite-momentum) spin waves. The rotor angular momentum ${\bm K}$ can be thought of as the projection of the collective spin of the system ${\bm J}$ on the Dicke-state sector of maximum square modulus ${\bm J}^2 = NS(NS+1)$; while spin waves describe a (weak) leakage of the dynamics away from this sector. The rotor dynamics is identical to that of a one-axis-twisting (OAT) model; RSW theory for the XXZ model is therefore justified when the dynamics of the complete model remains close to the OAT one, namely when $\langle \bm J^2 \rangle/[NS(NS+1)]$ remains close to unity during the evolution.  

RSW theory proceeds then as follows: it solves for the dynamics of the rotor and for that of the spin waves separately; and reconstructs all observables as stemming from a rotor and a spin-wave contribution, neglecting the same order of terms (quartic and further terms involving finite-momentum bosons) as those neglected in the Hamiltonian \cite{Roscildeetal2023, Roscildeetal2023b}. In so doing, the average magnetization takes the simple, exactly additive structure $\langle J^x \rangle = \langle K^x \rangle - N_{\rm bos}$; while the fluctuations of the collective spin components in the $yz$ plane transverse to the average magnetization stem uniquely from the rotor, ${\rm Var}(J^\perp) \approx {\rm Var}(K^\perp)$. The spin-squeezing parameter is therefore approximated as $\xi_R^2 \approx N{\rm Var}(K^{\rm min})/(\langle K^x \rangle - N_{\rm bos})^2$.

\begin{center}
\begin{figure*}[ht!]
\includegraphics[width=\textwidth]{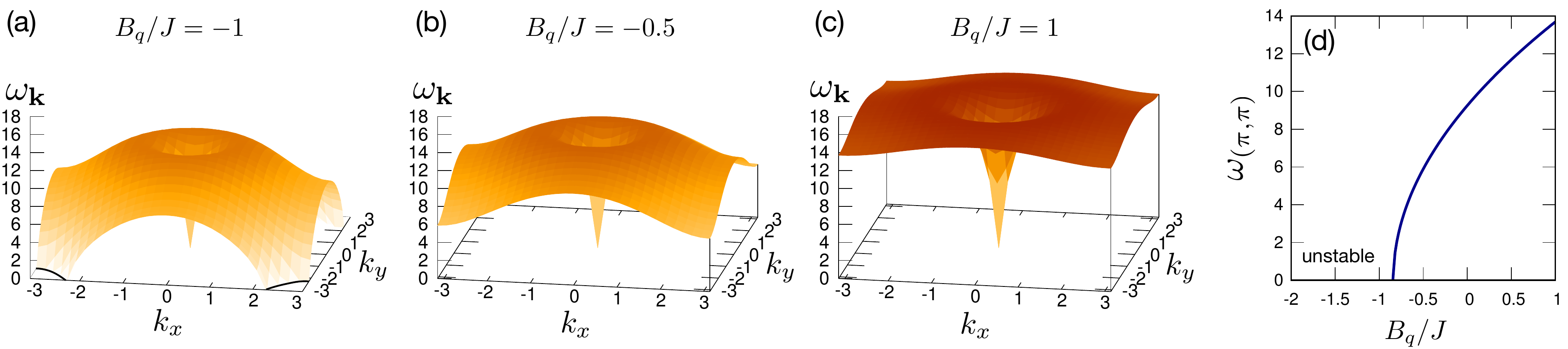}
\caption{Spin-wave spectrum of the 2$d$ dipolar XXZ model. (a-c) Spin-wave dispersion relations for $B_q=-J$, $-0.5 J$ and $J$. The black lines in panel (a) mark the onset of unstable (imaginary-frequency) modes. (d) Frequency of the mode at ${\bm k}=(\pi,\pi)$ as a function of $B_q$.}
\label{f.dispersion}
\end{figure*}
\end{center}

\section{Spin-wave spectrum for the dipolar XXZ model}
\label{s.SW}

The spin-wave dispersion relation is given by $\omega_{\bm k} = \sqrt{{\cal A}_{\bm k}^2 - {\cal B}_{\bm k}^2}$, where ${\cal A}_{\bm k}$ and ${\cal B}_{\bm k}$ are defined in Eq.~\eqref{e.AkBk} \cite{Roscildeetal2023, Roscildeetal2023b}. 
As shown in Fig.~\ref{f.dispersion} the dispersion relation has always a Goldstone mode at ${\bm k} = (0,0)$; but it also develops unstable modes, namely modes with imaginary frequency, at $\bm k = (\pi,\pi)$ and in its vicinity for $B_q < B_{q,m} \approx -0.845 J$. Within spin-wave theory, this result is independent of the spin $S$, which enters in the dispersion relation uniquely as a prefactor. This result signals that the assumption of long-range $xy$ ferromagnetism in the ground state breaks down for sufficiently negative $B_q$ values, in favor of a different form of spin order with pitch vector $(\pi,\pi)$ -- which corresponds to long-range N\'eel antiferromagnetism along the $z$ axis, as we shall further see in Sec.~\ref{s.QMC}. 
The instability of the spin-wave modes for $B_q < B_{q,m}$ leads to a rapid depolarization of the collective spin, and therefore to the breakdown of scalable spin squeezing. For this reason we have focused on positive values of $B_q$ in the study of spin-squeezing dynamics. 

Upon increasing $B_q$, the spin-wave dispersion relation does not develop any instability; yet the RSW dynamics is affected by an ever increasing $B_q$, as one can understand from Eqs.~\eqref{e.SMHam} and \eqref{e.AkBk}. Indeed ${\cal B}_{\bm k}$ represents the rate of production of pairs of HP bosons, and it grows linearly with $B_q$, leading to an ever faster and more significant proliferation of finite-momentum bosons. This aspect eventually leads to the breakdown of the assumption of diluteness of HP bosons upon which the RSW approach rests. 
$B_q$ enters as well in the expression of the inverse moment of inertia of the rotor Eq.~\eqref{e.inertia}, so that its growth accelerates the rotor dynamics. Nonetheless it is worth mentioning that, although RSW theory overestimates squeezing at large $B_q$ (as shown in the main text) the acceleration of spin-wave proliferation appears to overcome that of the rotor dynamics, and for large $B_q$ even RSW theory predicts the disruption of OAT-like squeezing dynamics, albeit at much larger $B_q$ values than what observed in our TCE calculations. 

\section{Truncated cumulant expansion for large-$S$ spins}
\label{s.TCE}

As explained in the main text, a sufficiently strong quadratic Zeeman shift $B_q$ can lead to a strong decrease of the collective spin length, bringing therefore the dynamics away from the OAT-like regime to which RSW theory applies. In order to describe the dynamics away from this regime, we adopt a fully alternative approach to RSW theory, based on a truncated cumulant expansion (TCE) of the multivariate quantum fluctuations of the spin ensemble. In general, given local observables $A_i$ defined on sites $i$, the cumulants are defined recursively as
\begin{align}
& \langle A_i \rangle  =  \langle A_i \rangle_c \\
&\langle  A_i A_j \rangle  =  \langle A_i A_j \rangle_c + \langle A_i \rangle_c \langle A_j \rangle_c \\
&\langle  A_i A_j A_k \rangle  =  \langle A_i A_j  A_k\rangle_c + \langle A_i A_j  \rangle_c \langle A_k\rangle_c  \\ 
&+ \langle A_i A_k  \rangle_c \langle A_j\rangle_c + \langle A_j A_k  \rangle_c \langle A_i\rangle_c +   \langle A_i \rangle_c \langle A_j \rangle_c \langle A_k \rangle_c
\nonumber \\
& (...) \nonumber
\end{align}
In our approach we aim at describing accurately the behavior of single sites -- so as to account correctly for the quadratic Zeeman shift in the regime in which it dominates the short-time dynamics. To this scope, we consider a complete set of on-site operators, whose linear superpositions span all possible local observables. Considering \emph{e.g.} the basis of local $S_i^z$ eigenstates $S_i^z |i;m_s\rangle = m_s  |i;m_s\rangle$, these operators are readily identified as 
\begin{equation}
T_i^{m_s,m_s'} = |i;m_s\rangle \langle i; m_s'|
\end{equation}
such that $A_i = \sum_{m_s, m_s'} \langle i; m_s|A_i | i; m_s'\rangle~ T_i^{m_s,m_s'}$ for any local operator $A_i$. 

In order to describe exactly the dynamics of a system of $N$ sites, one should then reconstruct the evolution of all correlation functions of the $T$ operators, $\{ \langle T_i^{m_s,m_s'} \rangle \}$,   $\{ \langle T_i^{m_s,m_s'} T_j^{n_s,n_s'} \rangle \} $, $\{ \langle T_i^{m_s,m_s'} T_j^{n_s,n_s'} T_k^{p_s,p_s'} \rangle \}$, etc. or alternatively of each cumulant, up to $N$-th order correlation functions or cumulants. In practice the strategy one can follow is to assume that the cumulants are structured in a hierarchy, such that higher-order cumulants are smaller than lower-order ones, as it is the case in many relevant physical examples (e.g. for bosonic or fermionic gases in states close to Gaussian states). If such a hierarchy exists, it can be conveniently truncated at some finite order, assuming that all $n$-th order cumulants for $n>n_0$ vanish -- this defines then a $n_0$-th-order TCE. If one takes $n_0=1$, the TCE approach amounts to a single-site mean-field approximation. Since our goal is to describe the dynamics of quantum correlations, we adopt instead a 2nd-order TCE, assuming the vanishing of all cumulants starting from 3rd order. 
Under this assumption, the evolution of the state can be fully described by knowing the evolution of the expectation values $\{ \langle  T_i^{m_s,m_s'} \rangle \} $ and $\{ \langle  T_i^{m_s,m_s'} T_j^{n_s,n_s'}  \rangle \}$, namely a number of objects $\sim O((2S+1)^4 N)$ for a translationally invariant system. It is therefore apparent that the computational cost increases substantially upon increasing the spin length $S$, so that the system sizes accessible in practice to TCE decrease with $S$.  

Considering the expectation values $\langle O \rangle \in \{ \langle T_i^{m_s,m_s'} \rangle , \langle T_i^{m_s,m_s'} T_j^{n_s,n_s'} \rangle \} $, their evolution is dictated by the Heisenberg equation 
\begin{equation}
i\frac{d\langle O \rangle}{dt} = \langle [ O, {\cal H} ] \rangle~. 
\end{equation}
The truncation of the cumulant hierarchy allows then to write the expectation value of the commutator on the right-hand side as a function of the first- and second-order cumulants for the $T$ operators, namely $\langle [ O, {\cal H} ] \rangle = F_O(  \{ \langle T_i^{m_s,m_s'} \rangle , \langle T_i^{m_s,m_s'} T_j^{n_s,n_s'} \rangle \} ) $, so that the equations of motion take the form of a closed set of non-linear differential equations.

\begin{center}
\begin{figure}[ht!]
\includegraphics[width=0.9\columnwidth]{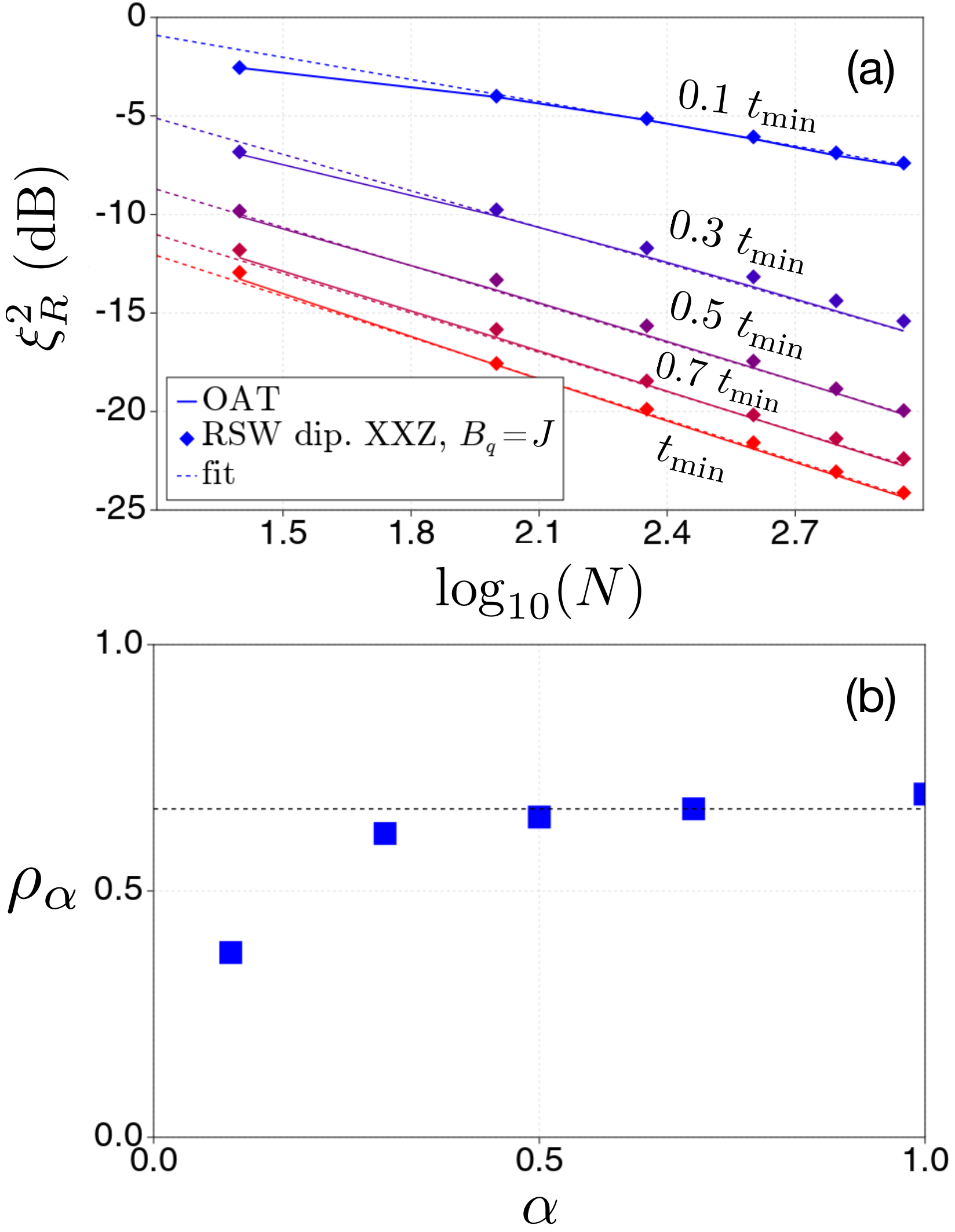}
\caption{Scaling of squeezing at times earlier than the optimal one in the one-axist-twisting model. (a) Scaling of the squeezing parameter at times $t_\alpha = \alpha t_{\rm min}$ with $\alpha =$ 0.1, 0.3, 0.5, 0.7 and 1. We compare the predictions of the OAT model for $N$ spins of length $S=3$ with those of RSW theory for the 2$d$ dipolar XXZ model with $B_q/J =1$ and the same system size. (b) Exponent $\rho_\alpha$ of the power-law scaling $\xi_{R}^2(t_\alpha) \sim N^{-\rho_\alpha}$, fitted to the OAT results as in the previous panel.}
\label{f.earlier_times}
\end{figure}
\end{center}

Writing the equations of motion of the first- and second-order correlation functions leads to very lengthy expressions which can nonetheless be obtained in a rather straightforward manner -- therefore we shall not report them here. We solve the coupled  $\sim O((2S+1)^4 N)$ non-linear equations of motion via a standard RK4 scheme. In particular the expectation value of the Hamiltonian can be written as a linear combination of first- and second-order correlation functions, so that the energy is conserved to machine precision by this approach. Notice moreover that the TCE approach involves uniquely an approximation at the level of the inter-site correlations, but no approximation is made instead on the statistics of fluctuations of single sites, whose physics is therefore described exactly -- in fact even the two-site physics is exactly described with a 2nd-order TCE. 

In spite of its ability to correctly describe single-site and two-site physics, the TCE approach has the fundamental shortcoming that no physical quantum state of the spin ensemble may in fact verify the hypothesis of vanishing cumulants beyond a given order $n_0 \geq 2$. Indeed, while Gaussian states (with exactly vanishing third- and higher-order cumulants) exist for ensembles of bosonic and fermionic modes, no such states are known for spin ensembles. As a consequence the TCE approach can produce unphysical results, especially so when correlations become strongly nonlocal. For the dynamics we investigated, the initial state has vanishing cumulants beyond first order, so that it is exactly described by 2nd-order TCE, as well as its short-time evolution. Yet when the dynamics of the 2d array of dipolar spins leads to the appearance of very strong squeezing -- specifically, when it is akin to that of the OAT model -- we observe that the Heisenberg inequality for the collective spin $ {\rm Var}(J^{\rm min}) {\rm Var}(J^{\rm max}) \geq \langle J^x \rangle^2/4$ can be violated by our TCE results, where $J^{\rm max}$ is the anti-squeezed collective-spin component, perpendicular to $J^{\rm min}$ in the $yz$ plane. This violation is the precursor to a drastic decrease of ${\rm Var}(J^{\rm min})$, which can become negative -- underlying the fact that the 2nd-order TCE approach has the tendency to strongly overestimate the anti-correlations associated with squeezing. In this situation we monitor the evolution of the ratio  $R = 4 {\rm Var}(J^{\rm min}) {\rm Var}(J^{\rm max}) / \langle J^x \rangle^2$, and we stop the time evolution when $R(t)$ reaches a maximum, which is the prelude to the rapid fall to values below unity and even negative. This is the reason why the TCE results shown in Fig.~1 of the main text are restricted to short times.

\begin{center}
\begin{figure*}[p!]
\includegraphics[width=\textwidth]{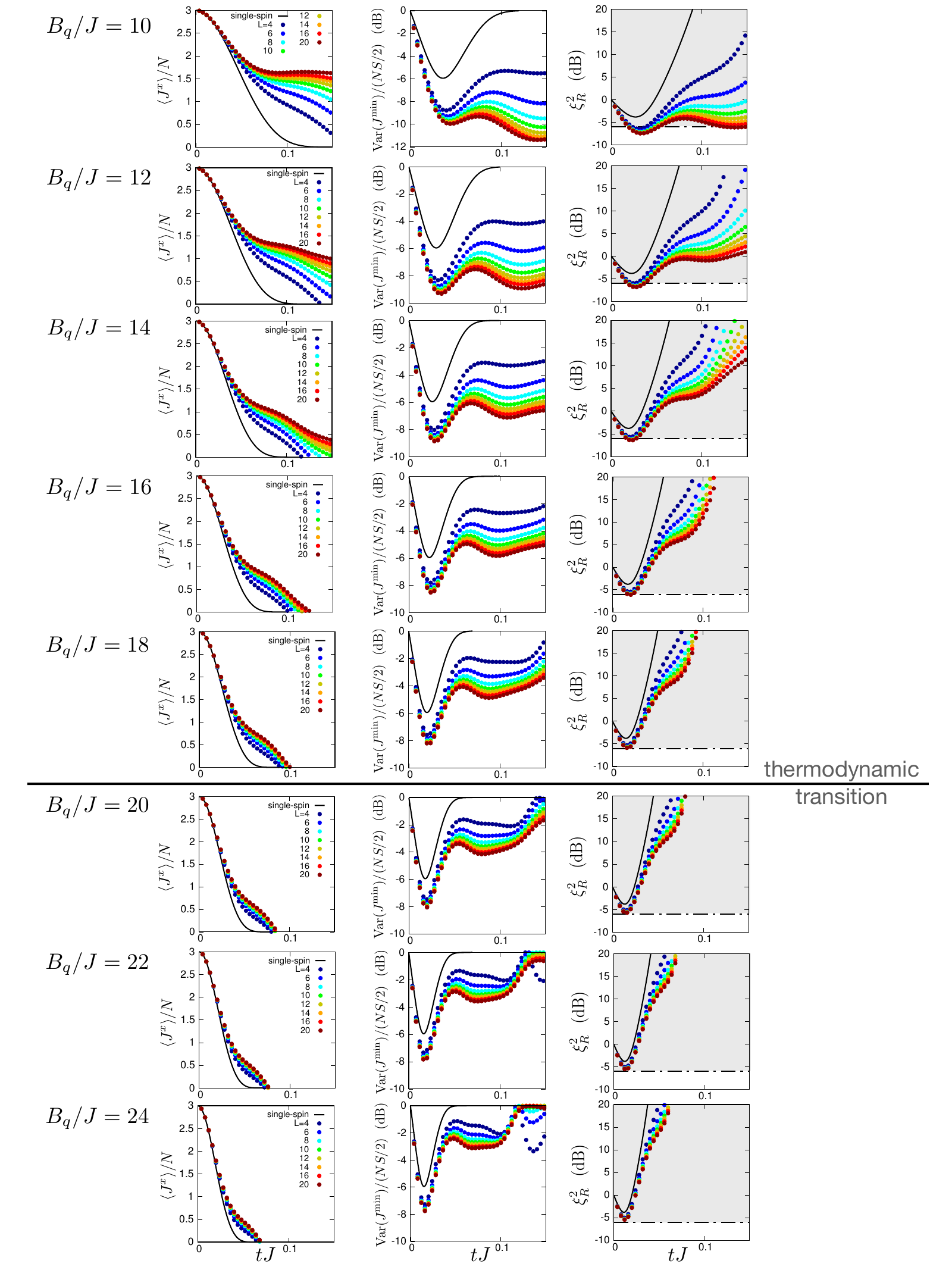}
\caption{Extended data for the dynamics of the 2d dipolar XXZ model with $S=3$ close to the thermodynamic transition driven by the quadratic Zeeman shift $B_q$. First column: average magnetization; second column: reduction of the minimum transverse variance with respect to the coherent-spin state value $NS/2$; third column: squeezing parameter.}
\label{f.transition_dynamics}
\end{figure*}
\end{center}

\section{Scaling of squeezing in the OAT model before the optimal time}
 
 In the case of an OAT-like spin squeezing dynamics, the unphysical results produced by the TCE approach prevent our calculations from reaching the optimal squeezing time of the dynamics; as a result we cannot study the scaling of the optimal squeezing from TCE calculations, and compare it with the RSW predictions. Nonetheless scalable squeezing in the OAT-like dynamics is not only a property of the squeezing parameter at the optimal time, but it is also present at earlier times, when one takes a time that scales with system size in the same way as the optimal one. In particular, for the system sizes we have considered, the optimal time of squeezing for an OAT model with spin length $NS$ is numerically found to scale as $t_{\rm min} = 2IA/(2NS)^\sigma$ where $A = 1.0142$ and $\sigma \approx 0.648$. In Fig.~\ref{f.earlier_times}(a) we show the squeezing parameter at earlier times $t_\alpha = \alpha t_{\rm min}$ with $\alpha = 0.1, 0.3, 0.5$ and 0.7. One can observe that a slower scaling of squeezing is obtained for these intermediate times, which is nonetheless a precursor to the scaling of the optimal squeezing. In particular we observe that the squeezing at earlier times $t_\alpha$ scales asymptotically as $\xi_R^2(t_\alpha) \sim N^{-\rho_\alpha}$, with an exponent $\rho_\alpha$ which approaches rather fast the value of the optimal squeezing $\rho_1 \approx 2/3$ (Fig.~\ref{f.earlier_times}(b)). The same behavior is found in the RSW predictions for the 2$d$ dipolar XXZ model for moderate $B_q$ values, as also shown in the figure.

\section{Extended data close to the thermodynamic transition} 

Fig.~\ref{f.transition_dynamics} shows the time dependence of the average magnetization per spin $\langle J^x \rangle/N$, of the minimal variance for the transverse-field components ${\rm Var}(J^{\rm min})$, and of the resulting spin squeezing parameter $\xi_R^2$ for the $S=3$ two-dimensional XXZ model, as obtained via the TCE approach described in Sec.~\ref{s.TCE}. We show data for various values of the quadratic Zeeman shift $B_q$ going across the thermodynamic transition from ferromagnetic to paramagnetic in the state at long times (see also the following section, Sec.~\ref{s.QMC}); and for each $B_q$ value we show data for various system sizes (from $4\times 4$ to $20 \times 20$).

\begin{center}
\begin{figure}[ht!]
\includegraphics[width=\columnwidth]{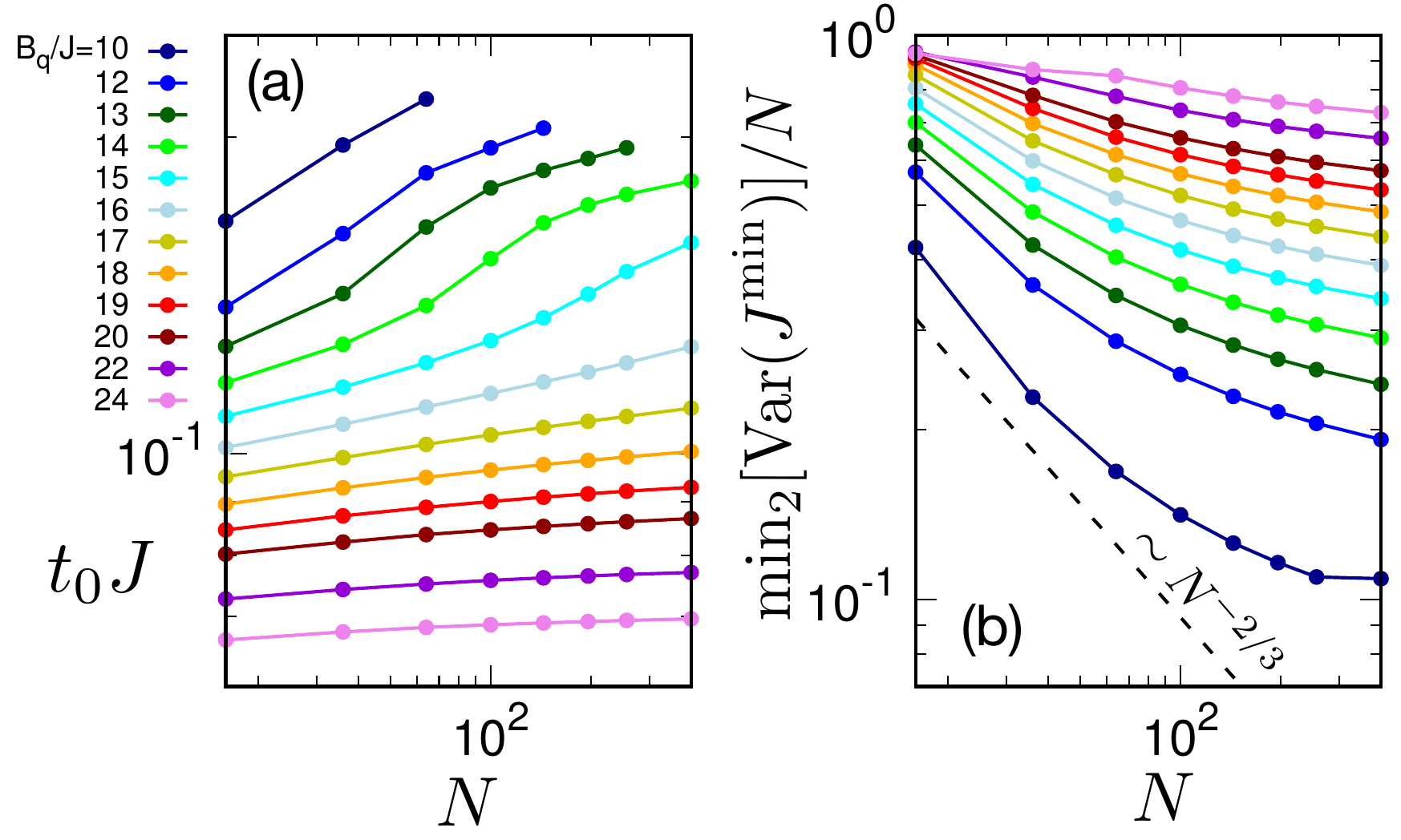}
\caption{Scaling of dynamical features from the TCE results for the $S=3$ dipolar XXZ model. (a) Time $t_0$ at which $\langle J^x\rangle$ crosses zero; (b) Scaling of the second minimum of ${\rm Var}(J^{\rm min})/N$ in time.}
\label{f.scaling_dynamics}
\end{figure}
\end{center}

We remark that in the regime in which the long-time behavior is expected to exhibit long-range ferromagnetism -- namely for $B_q \leq B_{q,c} \approx 18 J$ -- the TCE data exhibit the characteristic finite-size dynamical precursor of (infinite-size) spontaneous symmetry breaking, namely the persistence of a net magnetization $\langle J^x \rangle/N > 0$ for longer times the larger the system size. This persistence appears as a tail scaling with system size in the time dependence of $\langle J^x \rangle/N$ -- following an initial, size-independent drop, which is very well captured by the dynamics of single spins subject to the quadratic Zeeman field $B_q$. In particular we estimate the extent of this tail from the time $t_0$ at which $\langle J^x \rangle$ crosses zero. The scaling of this time with system size for various values of $B_q$ is shown in Fig.~\ref{f.scaling_dynamics}(a): we observe that the scaling slows down quite markedly for $B_q \approx B_{q,c}$, reflecting the thermodynamic transition. This is a signature that, in spite of their approximate nature, the TCE results appear to be sensitive to the transition itself, whose position is obtained instead from unbiased quantum Monte Carlo results (Sec.~\ref{s.QMC}).

Concomitantly with the strong reduction of scaling properties in the $\langle J^x \rangle/N$ at the thermodynamic transition, we observe that the dynamics of the minimal variance ${\rm Var}(J^{\rm min})$ maintains some significant scaling, not in its first minimum in time -- which is clearly dictated by the single-spin dynamics -- but rather in its second minimum. The depth of the second minimum $\min_2[ {\rm Var}(J^{\rm min})]$ is shown in Fig.~\ref{f.scaling_dynamics}(b) to exhibit a persistent scaling even past the transition. 
Nonetheless the scaling of the second minimum does not appear to follow a power law, and it is very different from the scaling (as $N^{-2/3}$ when considering the variance per spin) which is exhibited by the OAT model. Hence one might wonder whether this scaling is actually persistent asymptotically; and whether it can lead to scalable squeezing when $\langle J^x \rangle/N$ remains finite -- namely for $B_q < B_{q,c}$. A further possibility is that the scaling features of the dynamics involve an even later minimum in ${\rm Var}(J^{\rm min})$, which leads to a scaling minimum in the squeezing parameter observable only for much bigger system sizes and much longer evolution times than those explored here.

\begin{center}
\begin{figure}[ht!]
\includegraphics[width=0.8\columnwidth]{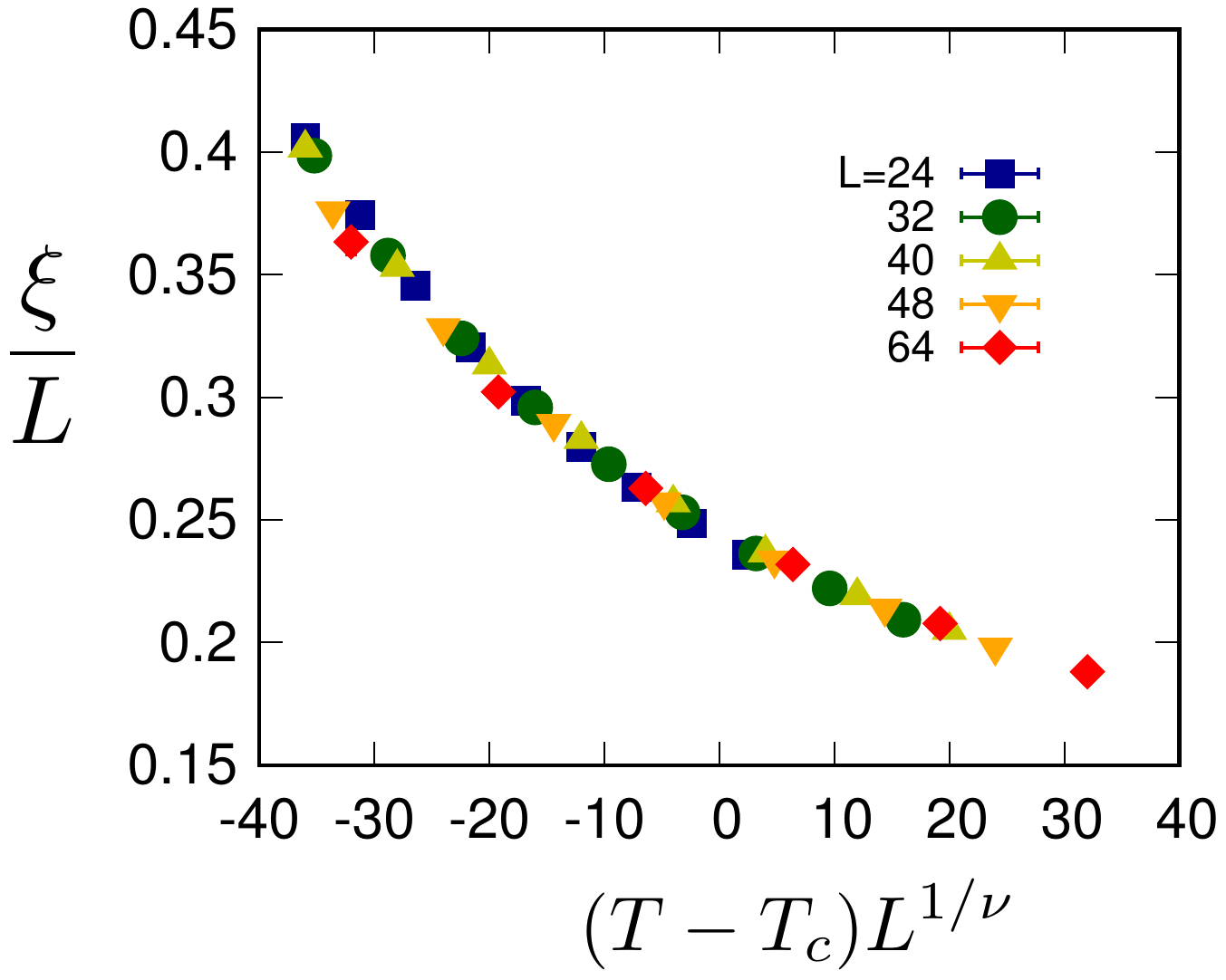}
\caption{Scaling plot of the correlation length at the ferromagnetic/paramagnetic transition for the 2d dipolar XXZ model with $S=3$ and $B_q/J=17$, using the critical exponent $\nu = 1$ and the critical temperature $T_c = 20.1 J$~. The data stem from unbiased QMC simulations. }
\label{f.scalingQMC}
\end{figure}
\end{center} 
 
\section{Equilibrium phase diagram from quantum Monte Carlo, and mean-field theory results for $S=6, 8$} 
\label{s.QMC}
 
 The QMC simulations have been performed using the Stochastic Series Expansion approach with directed-loop updates \cite{Syljuasen2002PRE}; this approach can be adapted to spin systems with arbitrary spin lengths $S$ by using \emph{e.g.} simple heat-bath transition probabilities dictating the directed-loop dynamics \cite{Aletetal2005}. We have performed simulations on $N = L \times L$ square lattices with periodic boundary conditions, and $L$ ranging from 24 to 64. The critical temperature $T_c$ for the ferromagnetic-paramagnetic transition of the dipolar XXZ model has been estimated via the scaling of the correlation length $\xi$, which can be extracted from the structure factor 
 \begin{equation}
 S(\bm q) = \frac{1}{N} \sum_{ij} e^{i\bm q \cdot (\bm r_i - \bm r_j)} \langle S_i^x S_j^x \rangle
  \end{equation}
  via the second-moment estimator
  \begin{equation}
  \xi = \frac{L}{2\pi} \sqrt{\frac{S(0,0)}{S(2\pi/L,0)} - 1}~.
  \end{equation}
  
  This quantity is expected to exhibit the following scaling behavior at the 2d dipolar transition 
  \begin{equation}
  \xi \sim L ~F_\xi(|T-T_c|L^{1/\nu})
  \end{equation} 
  where $F_\xi$ is a universal scaling function,  and $\nu = 1$ \cite{Defenuetal2023}. A representative scaling plot is shown in Fig.~\ref{f.scalingQMC}, showing a very good collapse of the QMC data. 

We have also investigated the thermodynamics of the large-$S$ dipolar XXZ model on a square lattice making use of mean-field theory, which postulates that the thermal state of the system can be written as a separable state $\rho(T) = \otimes_{i=1}^N \rho_i(T)$, where 
\begin{equation}
\rho_i(T) = \frac{\exp(-\beta {\cal H}_i)}{{\rm Tr}[\exp(-\beta {\cal H}_i)]} ~.
\end{equation}
${\cal H}$ is the effective single-site Hamiltonian
\begin{equation}
{\cal H}_i =  - {\bm H}_i \cdot ({\bm S}_i - \langle \bm S_i \rangle/2)  + B_q (S_i^z)^2
\end{equation}
containing the effective self-consistent field 
\begin{eqnarray}
H_i^{x(y)} & = & \frac{J}{2} \sum_{j(\neq i)} D_{ij} \langle S_j^{x(y)} \rangle \nonumber \\
H_i^{z} & = & - J \sum_{j(\neq i)} D_{ij} \langle S_j^{z} \rangle~
\end{eqnarray}
where $\langle S_i^\mu \rangle = {\rm Tr}(S_i^\mu \rho_i)$. 
The equilibrium state of the system is then evaluated by self-consistently calculating the average spin $\langle \bm S_i \rangle$ until convergence. In particular in the $xy$-ferromagnetic phase the average spin points in the $xy$ plane; it vanishes in the paramagnetic phase (including the one at $T=0$ for large $B_q$); while it is staggered along the $z$ axis in the $z$-N\'eel phase obtained for $B_q < B_m$. 

Fig.~3 of the main text shows that the phase diagram reconstructed via mean-field (MF) theory in the case $S=3$ is in relatively good agreement with the one obtained via QMC, even though the nature of the transition is not correct (\emph{e.g.} MF theory predicts $\nu=1/2$ instead of $\nu =1$). The quantitative accuracy is not surprising, given the fact that MF theory becomes exact in infinite dimensions;  and that the long-range tail of dipolar interactions in 2$d$ dominates the thermodynamics of the system, leading to the existence of a finite-temperature transition. 

\begin{center}
\begin{figure}[ht!]
\includegraphics[width=\columnwidth]{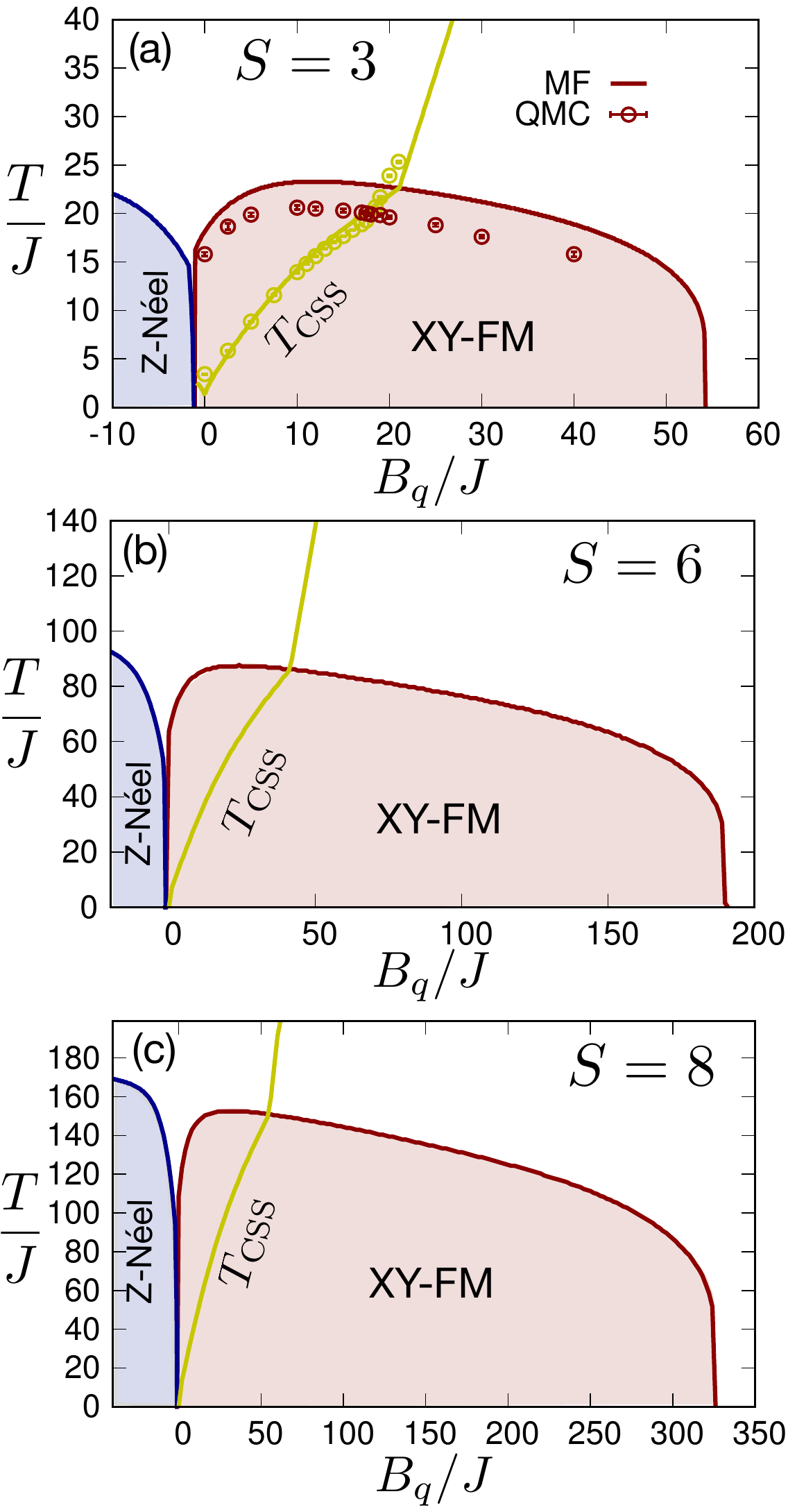}
\caption{Phase diagrams for the 2$d$ dipolar XXZ model with $S=3$, 6 and 8 from QMC and MF theory.}
\label{f.MF}
\end{figure}
\end{center}

Given the relatively good accuracy of MF theory in reconstructing the phase diagram, it can be used also for the cases $S=6$ and $S=8$, which are much more demanding for unbiased QMC calculations. Fig.~\ref{f.MF} compares the $S=3$ phase diagram with those for $S=6$ and $S=8$. MF theory predicts a thermodynamic transition at the $T_{\rm CSS}$ temperature for $B_{q,c} \approx 21 J$ for $S=3$ (to be compared with $B_{q,c} \approx 18 J$ from QMC); $B_{q,c} \approx 41 J$ for $S=6$, and  $B_{q,c} \approx 54 J$ for $S=8$. As we can see, $B_{q,c}$ appears to grow linearly with $S$. MF theory also allows one to predict the critical $B_q$ values delimiting the phase which exhibits $xy$ ferromagnetism. The transition to  N\'eel antiferromagnetism is predicted by MF theory to occur for $B_{q,m} \approx  -1.1 J$ for $S=3$, $B_{q,m} \approx  - J$ for $S=6$ and $B_{q,m} \approx  -0.95 J$ for $S=8$. These predictions are to be compared with $B_{q,m} \approx -0.845 J$ (for all spin values) from spin-wave theory; MF and spin-wave theory, albeit not predicting the same critical field, agree on its weak (or absent) spin-length dependence. On the other hand, MF theory predicts a transition from $xy$ ferromagnetism to quantum paramagnetism in the ground state for $B_q> B_{q,p}$ with $B_{q,p} \approx 54 J$ for $S=3$,  $B_{q,p} \approx 190 J$ for $S=6$ and $B_{q,p} \approx 326 J$ for $S=8$. The $B_{q,p}$ values appear to grow with $S$ approximately as $S^2$. As a consequence the range of $B_q$ values for which one observes $xy$ ferromagnetism at the CSS temperature $T_{\rm CSS}$ appears to shrink with respect to the whole $B_q$ extent of the $xy$ ferromagnetic phase, due to the different scaling of $B_{q,c}$ and $B_{q,p}$ with $S$.

\bibliography{largeS.bib}

\end{document}